\documentclass[a4paper]{article}

\usepackage{INTERSPEECH2019}
\usepackage{url}
\usepackage{xcolor}
\usepackage{multirow}

\makeatletter
\g@addto@macro{\UrlBreaks}{%
\do\/\do\-\do\_}
\makeatother

\title{LEAP Submission to CHiME-6 ASR Challenge}
\name{Anirudh Sreeram,
      Anurenjan Purushothaman,
      Rohit Kumar,
      Sriram Ganapathy.
    }

\address{
 Learning and Extraction of Acoustic Patterns (LEAP) lab \\
  Indian Institute of Science, Bangalore, 560012.}
\email{Email:\{sanirudh, anurenjanr, rohitk, sriramg\}@iisc.ac.in}

\begin{document}

\maketitle

% TODO
% - either intro or conclusion, say something about number of registrants for DIHARD II: 48

% Abstract
%
% - MAX LENGTH: 200 words
% - MUST match abstract on START
\begin{abstract}
    This paper reports the LEAP submission to the CHiME-6 challenge. The CHiME-6 Automatic Speech Recognition (ASR) challenge Track 1 involved the recognition of speech in noisy and reverberant acoustic conditions in home environments with multiple-party interactions. For the challenge submission, the LEAP system used extensive data augmentation and a factorized time-delay neural network (TDNN) architecture. We also explored a neural architecture that interleaved the  TDNN layers with LSTM layers. The submitted system improved the Kaldi recipe by 2\% in terms of relative word-error-rate improvements. 
    
\end{abstract}

\section{Introduction}
Automatic speech recognition (ASR) systems find widespread use in applications like human-machine interface, virtual assistants, smart speakers etc, where the input speech is often reverberant and noisy. the degradation of the systems in presence of noise and reverberation continues to be a challenging problem due to the low signal to noise ratio \cite{hain2012transcribing}.  For \textit{e.g.} Peddinti \textit{et al,} \cite{peddinti2017low} reports a $75\%$ rel. increase in word error rate (WER) when signals from a far-field array microphone are used in place of those from headset microphones in the ASR systems, both during training and testing. This degradation could be primarily attributed to reverberation artifacts \cite{yoshioka2012making, kinoshita2013reverb}. The availability of multi-channel signals can be leveraged for alleviating these issues as most of the real life far-field speech recordings are captured by a microphone array.

The presence of noise and reverberation in the far field speech signal degrades the performance of the ASR leading to increased Word Error Rates (WER). Many methods have focused on alleviating the degradation in the WER by using multi-channel microphones for far field speech recognition. The  traditional approach to multi-channel far-field ASR combines all the available channels by beamforming \cite{anguera2007acoustic} and then processes the resulting single channel signal effectively. Weighted Prediction Error (WPE)~\cite{wpe} is another classical signal processing technique used to remove reverberations from the far field signals. 

For automatic speech recognition acoustic modeling, the most popular modeling approach is the use of time delay neural networks \cite{peddinti2017low} that is trained with a lattice free maximum mutual information (MMI) criterion~\cite{povey2016purely}. The lattice free MMI model enable sequence cost function training and the model performs an efficient approximation to simplify computations without any lattice construction. Further, another powerful tool for dealing with noisy and reverberant speech in ASR is the use of  data augmentation \cite{ko2017study}.

The CHiME-6 represents the  sixth version in a series of challenges attempting to address automatic speech recognition in realistic home environments \cite{chime6_eval_plan}. The speech material was elicited using a dinner party scenario with efforts taken to capture data that is representative of natural conversational speech. This paper summarizes the LEAP submission to the Track-1 of the CHiME-6 challenge.

In our submission, we extend the previous literature of using data augmentation with a factorized version of time-delay neural network~\cite{povey2018semi}. In addition, we also explore the use of LSTM models in the factorized TDNN. The system description is given below followed by the section which reports the ASR results in the CHiME-6 development data. 

\section{The  CHiME-6 challenge dataset}
The dataset for CHiME-6 challenge is same as the dataset for CHiME-5 challenge. The dataset is made up of the recording of $20$ separate dinner parties taking place in real homes. Recordings were made in kitchen, dining and living room areas with eachphase lasting for a minimum of $30$ mins. Each dinner party has $4$ participants. Each party has been recorded with a set of $6$ Microsoft Kinect devices and in-ear Soundman OKM II Classic Studio binaural microphones. Each  Kinect  device  has  a  linear  array  of  $4$  sample-synchronised microphones. The data is split into training, development, and evaluation sets as follows.

The training set consist of $16$ parties with $32$ speakers in total. The number of utterances in the training set is $79,980$ adding up to around $41$ hours. Development set has $2$ parties with $8$ speakers and $7,440$ utterances with nearly $4.5$ hours of audio. Similarly $2$ parties with $8$ speakers and $11,028$ utterances comprising of $5.1$ hours of audio is used as the evaluation set. 
\begin{figure*}[t!]
    \centering
    \includegraphics[width=\textwidth,height=6.2cm]{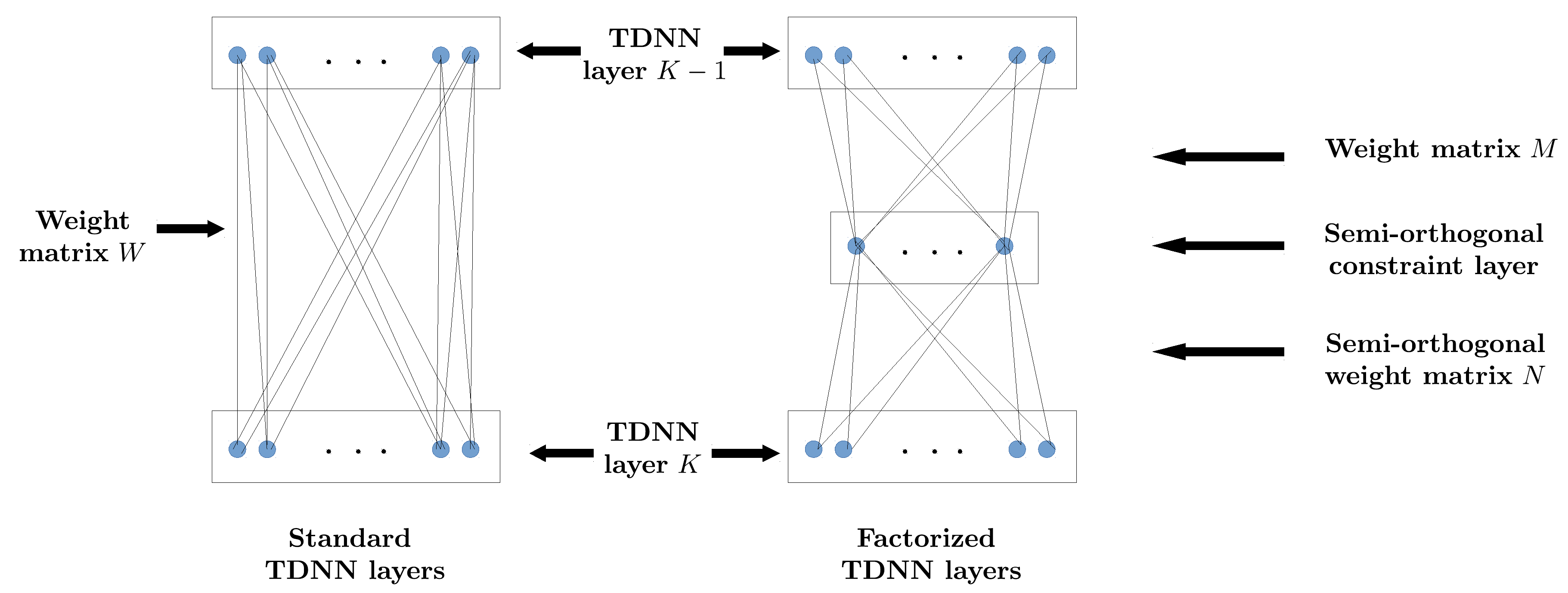}
    \caption{Standard TDNN and TDNN layers with semi-orthogonal weight constraint.}
    \label{fig:tdnnf}
\end{figure*}

\begin{table*}[t!]
\caption{Various model architectures and Word Error Rates (WER) \% results for the CHiME-6 development set.}
\begin{tabular}{@{}c|ccccccc@{}}
\toprule
\textbf{Models (No. of layers)}                            & \textbf{S02 DINING}  & \textbf{S02 Kitchen} & \textbf{S02 Living}  & \textbf{S09 Dining}  & \textbf{S09 Kitchen} & \textbf{S09 Living}  & \textbf{Overall}     \\ \midrule
\textbf{Kaldi-Recipe-Results~\cite{chime6_eval_plan}}                           & 53.80                     &56.47                      & 47.78                     & 53.76                     & 50.30                     & 49.92                     & 51.75         \\
\hline 
\textbf{HMM-GMM}                           & 84.39                     & 86.12                     & 78.63                      & 86.00                     & 83.67                      & 84.17                     & 83.27                     \\
\textbf{F-TDNN (15)}                   & 55.40                & 57.78                & 48.49                & 54.92                & 50.95                & 50.95                & 52.79                \\
\textbf{F-TDNN (18) }                   & 53.53                & 55.50                & 46.64                & 53.45                & 50.31                & 48.93                & \textbf{51.04}              \\
\multicolumn{1}{l|}{\textbf{F-TDNN (18)  LSTM (3)}} & 56.80 & 59.79 & 49.91 & 57.68 & 52.65 & 53.42 & 54.65 \\ \bottomrule
\end{tabular}
\label{tab:asr_wer}

\vspace{+0.1cm}

\end{table*}
\section{Proposed system}
% \subsection{Weighted Prediction Error (WPE) based speech enhancement}
% WPE based speech enhancement method is a statistical model-based speech dereverberation approach that can cancel the late reverberation of a reverberant speech signal captured by distant microphones without prior knowledge of the room impulse responses.
% \subsubsection{Observed Signal Model}
% \begin{equation}
%     x_t^{\small(m\small)} = \sum_{k=0}^{L_h-1}h^{\small(m\small)}_ks_{t-k} + b_t^{\small(m\small)}
% \end{equation}
% Where $s_t$, $x_t^{\small(m\small)}$ and $b_t^{\small(m\small)}$ are digitized sequences of the source, observed,
% and noise signals, respectively, where $t$ and $m$ are the sample
% and microphone (channel) indices, respectively. The room impulse
% response (RIR) of length $L_h$ from the source to the $m^{th}$ 
% microphone is denoted by $\bar{h}^{\small(m\small)}$ = $[h^{\small(m\small)}_0, h^{\small(m\small)}_1,...,h^{\small(m\small)}_{L_h-1}]^T$. In WPE based speech enhancement, we estimate a dereverberation filter that produces an enhanced signal $y_t$ that contains less reverberation than the observed signal $x_t^{\small(m\small)}$ without greatly increasing the acoustic noise level. The
% enhanced signal $y_t$ is given by,
% \begin{equation}
% y_t = \sum_{m=1}^{L_m} \sum_{k=0}^{L_w-1} w_k^{\small(m\small)}x_{t-k}^{\small(m\small)}
% \end{equation}
% where $\bar{w}_t^{\small(m\small)}$ = $[w^{\small(m\small)}_0, w^{\small(m\small)}_1,...,w^{\small(m\small)}_{L_w-1}]^T$ is the dereverberation filter of length $L_w$.

The proposed system consists of a pre-processing step using guided source separation based beamforming. Acoustic Model consists of an $18$ layer facorized time delay neural network (F-TDNN) trained with lattice free MMI cost function with L2 weight regularization and $15$-fold data augumentation.

%\subsection{Beamforming}
%Beamforming is the process of combining all the available microphone channels of information to get an improved single channel signal in terms of signal to noise ratio (SNR). In \cite{anguera2007acoustic}, a fully acoustic beamforming technique is suggested. A generalization of  the  well  known  weighted-delay and sum  beamforming  technique is used.  The signal output $z_t$ is expressed as the weighted sum of the different channels as follows,
%\begin{equation}
%    z_t = \sum_{m=1}^{M}W^m_ty^m_{t-TDOA^{(m, ref)_t}}
%\end{equation}
%where, $W_t^m$ is the relative weight for microphone $m$ (out of $M$ microphones)  at  instant $t$,  with  the  sum  of  all  weights equals   to   1, $y_t^m$ is   the   signal   for each   microphone channel and $TDOA^{(m, ref)_t}$ (Time Delay Of Arrival) is the relative delay between  each  channel  and  the  reference channel.
\subsection{Guided Source Separation (GSS)}
In the CHiME-6 challenge, clean uncorrupted speech is not available for training a mask estimating neural network \cite{gev_dnn} based beamformer. Here, one has to resort to unsupervised mask estimation techniques, which, e.g., have been used in the context of Blind Source Separation (BSS) \cite{bss} employing spatial mixture models. In \cite{gss_chime,gss_vincent} the BSS approach is modified to make efficient use of the available time and speaker annotations provided with the challenge data.
The  GSS  outputs  time-frequency  masks, from these mask spatial covariance matrices are estimated, and from these matrices, in turn, the coefficients of the statistically optimum Minimum Variance Distortionless Response (MVDR) beamformer \cite{mvdr} are computed.
\subsection{Factorized Time Delay Neural Networks (F-TDNN)}
Recently factorized typologies of conventional networks such as TDNN are suggested \cite{povey2018semi} as an improvement over TDNN-LSTM architectures with faster decoding. The basic idea is to factorize the weight matrix $W$ of a TDNN layer into two matrices as $W = M \times N$. Here, $N$ is constrained to be semi-orthogonal (a non-square matrix with orthogonal rows or columns). Figure \ref{fig:tdnnf} shows this difference more apparently. In our case, we trained the model with F-TDNN architecture with different number of layers and with data augmentation. The F-TDNN experiments used hidden layers of size $1536$ with a bottleneck dimension of $160$. In addition, we also explore an architecture of F-TDNN with a LSTM layer interleaved after every 5 layers of F-TDNN architecture. In the F-TDNN-LSTM implementation, the hidden dimension of TDNN layers was kept at $1024$ with a bottleneck dimension of $160$. The LSTM layers had a recurrent projection of $512$ dimensions and non-recurrent projections of $512$ dimensions. All the F-TDNN models also used a L2 weight regularization and were trained with LF-MMI cost function \cite{povey2016purely}. 

\subsection{Data Augmentation}
All the results for the F-TDNN models used a data augmentation of artificial room reverberation using $5$ small and medium rooms ~\cite{ko2017study}. In addition, a 3-way speed perturbation was also employed yielding $15$ copies of the original training data.

\section{Experiments and results}
A mono-phone model is trained initially with MFCC features, using this a tri-phone model improved the WER. A tri-phone model with linear discriminant analysis (LDA) and maximum likelihood linear transform (MLLT) is trained, which improved the performance. Finally, speaker adaptive training (SAT) is performed further improving the WER and is reported in Table~\ref{tab:asr_wer} as HMM-GMM.

The WER results for various systems are reported in Table~\ref{tab:asr_wer}. As seen here, the WER results for F-TDNN model with $15$ layers improved drastically over the HMM-GMM framework. This is further improved by the model with $18$ layers. However, the inclusion of LSTM layers did not benefit the ASR performance. The final submitted system from the LEAP team was the F-TDNN ($18$ layer) model. This submitted system improved the Kaldi recipe by $2$\% in terms of relative word-error-rate improvements.

% \usepackage{booktabs}

% TODO: Remove following line.
\bibliographystyle{IEEEtran}
\bibliography{refs}

\end{document}